\documentclass[aps,pra,twocolumn,byrevtex,showpacs,floatfix,superscriptaddress,reprint]{revtex4-2}
\usepackage{graphicx}
\usepackage{subcaption}
\usepackage{float}
\usepackage{physics}
\usepackage{amsmath,amsfonts,amsthm,bm}\usepackage{epsfig}
\usepackage{hyperref}

\usepackage{caption}
\graphicspath{{pict/}{}}

\usepackage{mathrsfs}
\usepackage{mathtools}

\usepackage[ruled,vlined]{algorithm2e}

\hypersetup{colorlinks=true,linkcolor=blue,citecolor=blue,urlcolor=blue}

\begin{document}

\title{ Self-guided certification of nonlocality in quantum networks}

\author{J. Cort\'es-Vega}
\email[corresponding author: ]{jeancortes@udec.cl}
\affiliation{Instituto Milenio de Investigaci\'on en \'Optica, Universidad de Concepci\'on, Concepci\'on, Chile}
\affiliation{Facultad de Ciencias F\'isicas y Matem\'aticas, Departamento de F\'isica, Universidad de Concepci\'on, Concepci\'on, Chile}

\author{L. Pereira}
\affiliation{ICFO - Institut de Ciencies Fotoniques, The Barcelona Institute of Science and Technology, 08860 Castelldefels, Barcelona, Spain}

\author{A. Delgado}
\affiliation{Instituto Milenio de Investigaci\'on en \'Optica, Universidad de Concepci\'on, Concepci\'on, Chile}
\affiliation{Facultad de Ciencias F\'isicas y Matem\'aticas, Departamento de F\'isica, Universidad de Concepci\'on, Concepci\'on, Chile}

\date{\today}


\begin{abstract}

Bell's theorem shows that quantum theory is incompatible with local hidden-variable models. In recent years, research on nonlocality has moved beyond Bell's original scenario to quantum networks, where multiple independent sources distribute physical systems among distant parties, giving rise to correlations certified by nonlinear rather than standard Bell inequalities. Here, we introduce a self-guided protocol that variationally optimizes each party's measurement to maximize the violation of a network Bell inequality, with the violation evaluated efficiently at each step via local Pauli classical shadows and the search driven by the Complex Simultaneous Perturbation Stochastic Approximation (CSPSA) algorithm. Once converged, the measurement settings it returns are implemented directly and the inequality is re-evaluated without shadows. This two-stage structure separates a device-dependent search from a certificate that depends only on the observed statistics and on the causal structure of the network. We validate the protocol by numerically simulating it on the triangle network using the Wagon-Wheel inequality, recovering the violation achieved by the Fritz distribution, and extending the certification to non-maximally entangled and noisy states.

\end{abstract}

\maketitle

\section{Introduction}

Since its publication in 1964 \cite{Bell}, Bell's theorem has played a central role in the development of quantum information theory. It shows that no theory based on local hidden variables can fully explain the predictions of quantum mechanics \cite{brunner2014}. The theorem has influenced our understanding of concepts such as causality, entanglement, and even the nature of physical reality itself. Over the decades, experimental tests of Bell's ideas have continued to evolve. Today, its influence extends to modern quantum network architectures, where multiple independent sources and parties introduce even deeper challenges to our classical intuition \cite{tavakoli2021}.

Entanglement swapping is a fundamental quantum information protocol in which two particles that have never interacted become entangled via intermediate systems. In its standard form, two initially independent entangled pairs are prepared, and a joint measurement on one particle from each pair projects the remaining two spatially separated particles into an entangled state. This process illustrates the nonlocal correlations that arise not from direct interaction but from the structure of quantum measurements and shared entanglement \cite{zukowski1993}.

Building on these ideas, \textit{quantum networks} generalize the concept of entanglement swapping to more complex causal structures involving multiple parties and independent sources of entanglement. A prominent example is the triangle network, where three parties are pairwise connected by independent sources and no party shares a direct entangled state with all others. These scenarios extend the scope of Bell-type tests and enable the study of nonlocality in settings without measurement input, challenging our understanding of classical correlations and causal inference \cite{branciard2012, luo2024}.

Within this framework, the \textit{Fritz distribution} \cite{fritz2012} emerges as a key quantum distribution that violates the classical constraints imposed by the locality of the network. Constructed using maximally entangled states and specific local measurements, the Fritz distribution demonstrates that quantum mechanics allows correlations in network scenarios that cannot be explained by any classical causal model that obeys the independence of the sources. This result generalizes the notion of Bell nonlocality to networks and highlights the need for new tools, such as causal compatibility inequalities, to detect quantum effects in structured scenarios \cite{renou2019}.

\begin{figure}[t!]
    \centering
    \includegraphics[width=0.8\linewidth]{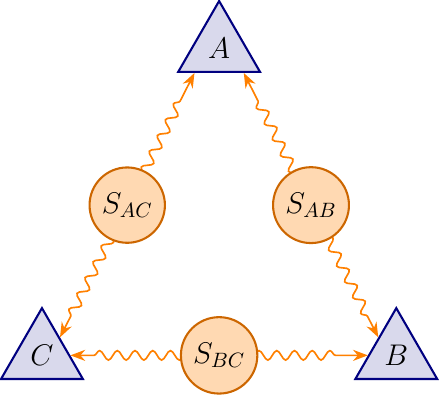}
    \caption{Cyclic structure of the triangular network. Three independent sources, $S_{AB}$, $S_{BC}$ and $S_{AC}$, distribute bipartite systems among nodes $A$, $B$ and $C$, in such a way that each node receives one system from each of the two sources it is connected to and performs a single joint measurement upon them.}
    \label{fig:triangle}
\end{figure}

The search for optimal quantum strategies in quantum networks is a challenging optimization problem because it involves non-linear objective functions \cite{rosset2016}. Classical approaches require full knowledge of the states and measurements and become computationally expensive as the size of the network increases \cite{liang2007,mortimer2025}. In addition, the non-linearity of the inequalities precludes the direct use of semidefinite programming \cite{tavakoli2024}. Recently, neural networks have been used to characterize genuine nonlocal correlations in Bell networks \cite{sunilkumar2025}. Self-guided algorithms offer a scalable approach to solving optimization problems in quantum information. They are founded on the ability to experimentally measure the objective function, which is optimized over a parametrized set of measurements by a gradient-free optimization algorithm such as SPSA \cite{spall1992}, CSPSA \cite{utreras2019}, or Nelder-Mead. This kind of optimizer enables efficient navigation of the optimization landscape with only noisy objective function evaluation, making it suitable for experimental scenarios \cite{chapman2016}. Self-guided quantum algorithms have been used to perform pure-state quantum state tomography \cite{ferrie2014,utreras2019,zambrano2020}, detect or estimate entanglement \cite{munozmoller2022,wang2022detecting,wang2021vqsvd}, and find the maximal violation of the CHSH inequality \cite{yang2017,poderini2022,cortesvega2023,dowling2025,tang2026}, among other applications \cite{gidi2023,gacon2021,concha2023}. 

In this work, we present a self-guided protocol to certify nonlocality of unknown quantum states in quantum networks by variationally maximizing nonlinear inequalities. The protocol has a two-stage structure: a device-dependent search and a direct certification. In the first step, we parameterize each measurement and maximize the value of the nonlinear inequality. The search is driven by CSPSA, a gradient-free stochastic optimization algorithm that uses two objective function evaluations per iteration to approximate the gradient. We employ classical shadows to estimate the objective function during the search, which makes the protocol practical even in settings where full state tomography is infeasible. The output of this search is used to certify nonlocality by a direct evaluation of the inequality, without resorting to classical shadows. Since the search does not require prior knowledge of the emitted states, the protocol is well suited to noisy experimental setups, such as next-generation quantum communication networks or distributed quantum computing. We demonstrate the performance of the protocol on the triangle network and the Wagon-Wheel inequality for sources that generate partially entangled and noisy states.

\section{Preliminaries}
In this section, we introduce the preliminary concepts for our proposal. First, we introduce the experimental scenario in which our protocol will be applied, the Bell network. Later, we introduce the CSPSA optimization algorithm and the classical shadow formalism as the main ingredients of our self-guided algorithm. 

\subsection{Quantum networks}

A quantum network consists of a set of $n$ parties $\mathcal{X}$ and a set of independent sources $\mathcal{S}$, where each source distributes a quantum state among a designated subset of parties. Each party collects the subsystems it receives and performs a joint measurement on them with inputs $\alpha=\{\alpha_x\}_{x\in\mathcal{X}}$, producing an output $a = \{a_x\}_{x\in\mathcal{X}}$. The network topology is encoded in a hypergraph, where the parties are vertices, and each source defines a hyperedge connecting the parties it serves. The key feature of this scenario is the independence of the sources, which imposes non-trivial causal constraints on the achievable distributions \cite{tavakoli2021}. The distribution $p(a|\alpha)$ is compatible with a classical hidden-variable model if and only if it admits a decomposition in the form
\begin{align}
    p(a|\alpha) = \int \prod_{s\in\mathcal{S}} {\rm d}\lambda_s \prod_{x\in\mathcal{X}} p_x(a_x| \alpha_x, \{\lambda_s\}_{s \rightarrow x}), \label{eq:hidden_model}
\end{align}
where $\{\lambda_s\}$ are independent local hidden variables associated with each source $s$, and each party $x$ receives only the variables from the sources connected to it. According to quantum mechanics, the probability distribution generated by a Bell network is
\begin{align}
    p(a|\alpha) = \Tr\left[\left(\bigotimes_{x\in\mathcal{X}} M_{a_x|\alpha_x}^{x}\right)\left(\bigotimes_{s\in \mathcal{S}} \rho_s\right)\right], \label{eq:quantum_model}
\end{align}
where $\{M_{a_x|\alpha_x}^{x}\}_{x\in\mathcal{X}}$ are positive-operator valued measures (POVM) for each party $x$ and $\rho_s$ is the state distributed by source $s$.

Detecting nonlocal correlations in quantum networks requires inequalities tailored to the specific network topology. Unlike the standard Bell scenario, the independence of the sources renders the set of classical correlations non-convex, so that linear inequalities are insufficient to characterize it \cite{dasilva2023}. As a consequence, the inequalities that capture $n$-local correlations are inherently nonlinear, and a violation of such inequalities certifies that no classical model compatible with the network structure can reproduce the observed correlations \cite{krivachy2020}. They have the following general form
\begin{align}
    I( p(a|\alpha) ) \leq I_{\rm Cl}, \label{eq:nonlocal_inq}
\end{align}
with $I( p(a|\alpha) ) $ a non-linear function of the probability distribution and $I_{\rm Cl}$ its classical limit. Iterative methods to construct such inequalities have been proposed for cyclic \cite{rosset2016} and tree-structured \cite{tavakoli2016} networks. These protocols start from a simpler network with a known Bell inequality, which is iteratively extended by adding one additional source and one additional party at each step, and then construct the corresponding inequality. A computationally efficient construction applicable to an arbitrary network has been proposed, based on expressing the inequalities as a maximal matching problem on a bipartite graph \cite{luo2018}. Beyond standard network nonlocality, a stronger notion of full network nonlocality certifies that all sources contribute with nonlocal resources \cite{pozaskerstjens2022}.

The triangle network, shown in Fig.\thinspace\ref{fig:triangle}, is a special case of a quantum network. It consists of three parties, Alice (A), Bob (B), and Charlie (C), pairwise connected via three independent bipartite sources $S_{AB}$, $S_{BC}$, and $S_{CA}$. Each source $S_{XY}$ distributes a bipartite quantum state $\rho_{XY}$ to the parties $X$ and $Y$, so that each party has two subsystems and performs a joint measurement on them, producing outputs $a$, $b$, and $c$, respectively. In contrast to the standard Bell scenario, there are no classical inputs in the measurement settings, so the joint probability distribution is simply $p(a,b,c)$.

Since each party performs a joint measurement on the two qubits it receives, its outcome takes one of four values, which we label with a bit pair: $a\in\{a_0,a_1,a_2,a_3\}$ corresponds to $(a_l,a_r)=(0,0),(0,1),(1,0),(1,1)$, respectively, and analogously for $b$ and $c$. Throughout, subindices indicate which bits a probability refers to, and the argument lists their values in the order in which the subindices appear. Thus $p_{A_lB_l}(10)$ is the marginal probability that $a_l=1$ and $b_l=0$, and $p_{C_lC_r}(01)$ the marginal probability that $c_l=0$ and $c_r=1$. A probability written without subindices and six-bit arguments denotes the full joint distribution, $p(a_la_rb_lb_rc_lc_r)  \equiv p(a,b,c),$ so that, for instance, $p(100110)$ is the probability of the outcome $(a_l,a_r)=(1,0)$, $(b_l,b_r)=(0,1)$, $(c_l,c_r)=(1,0)$.

Fritz demonstrated \cite{fritz2012} that the triangle network can produce quantum correlations that violate the classical constraint of Eq.\thinspace\eqref{eq:hidden_model}. In his construction, the source $Y$ distributes a two-qubit state that violates the CHSH inequality, with parties $A$ and $B$ measuring, on the qubit they receive from $Y$, the observable that maximizes this violation. The sources $X$ and $Z$ distribute maximally entangled states $\ket{\Phi^+}=(\ket{00}+\ket{11})/\sqrt{2}$, measured by the parties to which they connect ($A,C$ for $X$, $B,C$ for $Z$) on fixed computational basis $\{\ket{0},\ket{1}\}$. Under this choice, the bit obtained by $A$ from $X$ (denoted $a_l$) also determines which of the two optimal CHSH observables is measured on the qubit received from $Y$, and analogously for $B$ and $Z$. The party $C$, measuring both qubits on the fixed computational basis, then reports exactly those two bits as its own outcome, so that $c_l=a_l$ and $c_r=b_l$. In this construction, the bits $a_l$ and $b_l$ therefore play, for the CHSH test embedded in the triangle, the role that a measurement input plays in the standard Bell scenario. 

This perfect correlation between $C$ and the bits inherited from $A$ and $B$ collapses the support of the resulting distribution, denoted $p_{\rm F}$, from 64 to only 16 nonzero outcomes, organized in two blocks of eight according to the result obtained on the qubit shared through $Y$,
\begin{align}
    &p_{\rm F}(00,00,00)=p_{\rm F}(01,00,00)=p_{\rm F}(00,10,01) \nonumber\\
    &\quad=p_{\rm F}(01,10,01)=p_{\rm F}(10,01,10)=p_{\rm F}(11,01,10) \nonumber\\
    &\quad=p_{\rm F}(10,11,11)=p_{\rm F}(11,11,11) \nonumber\\
    &\quad=\frac{1}{32}\left(2+\sqrt{2}\right), \label{eq:fritz1}\\[4pt]
    &p_{\rm F}(00,01,00)=p_{\rm F}(01,01,00)=p_{\rm F}(00,11,01) \nonumber\\
    &\quad=p_{\rm F}(01,11,01)=p_{\rm F}(10,00,10)=p_{\rm F}(11,00,10) \nonumber\\
    &\quad=p_{\rm F}(10,10,11)=p_{\rm F}(11,10,11) \nonumber\\
    &\quad=\frac{1}{32}\left(2-\sqrt{2}\right), \label{eq:fritz2}
\end{align}
and every combination not listed above has zero probability.

By means of the Fritz distribution and the inflation technique, nonlocal inequalities for the triangle network can be constructed. The simplest of them is the Wagon-Wheel inequality \cite{fraser2018}, defined by the combination
\begin{align}
    I_{\rm WW} = \, & -p_{A_lB_l}(10) + p_{A_lB_lC_lC_r}(1010) \nonumber\\
    & - p_{A_lB_l}(01)p_{C_lC_r}(10) - p_{C_lC_r}(00)p_{C_lC_r}(11) \nonumber\\
    & - p_{C_lC_r}(01)p(100110) - p_{C_lC_r}(01)p(110010) \nonumber\\
    & + p_{C_lC_r}(00)p(101111) + p_{C_lC_r}(00)p(111011) \nonumber\\
    & + p_{C_lC_r}(10)p(001001) + p_{C_lC_r}(10)p(011101) \nonumber\\
    & + p_{C_lC_r}(11)p(000000) + p_{C_lC_r}(11)p(010100), \label{eq:wagonwheel}
\end{align}
which is quadratic in the probability distribution and has classical bound $I_{\rm Cl}=0$. When we evaluate this inequality on the Fritz distribution of Eqs.\thinspace\eqref{eq:fritz1}--\eqref{eq:fritz2}, we obtain
\begin{align}
    I_{\rm WW}^{\rm Fritz} = \frac{1}{16}\left(\sqrt{2}-1\right) \approx 0.0256,
\end{align}
which exceeds the classical bound $I_{\rm Cl}=0$ and certifies that the Fritz distribution is incompatible with the triangle-local model of Eq.\thinspace\eqref{eq:hidden_model}. We refer it as the \emph{Fritz value}. It is nevertheless the natural target for our protocol because we optimize the measurements at fixed states, and for three maximally entangled sources, the Fritz strategy is the best-known violation. 


\subsection{CSPSA algorithm}
The Complex Simultaneous Perturbation Stochastic Approximation (CSPSA) algorithm \cite{utreras2019} is an optimization method designed for real-valued functions of complex arguments of real value $f(z,z^*)$. Unlike traditional approaches that decompose complex parameters into their real and imaginary parts, CSPSA operates natively in the complex domain using the Wirtinger calculus \cite{kreutzdelgado2009}.

The iterative update rule of CSPSA is given by
\begin{equation}
	\hat{z}_{k+1} = \hat{z}_k + a_k \hat{g}_k(\hat{z}_k, \hat{z}_k^*),
	\label{eq:update}
\end{equation}
where $\hat{z}_k$ is the estimate at iteration $k$, $a_k$ is a positive learning rate, and $\hat{g}_k$ is an estimator of the Wirtinger gradient $\partial f/\partial z^*$. The plus sign in Eq.\thinspace\eqref{eq:update} corresponds to gradient ascent, which is the relevant case here since we maximize the violation. 

The gradient estimator is constructed as
\begin{equation}
	\hat{g}_{k,i} = \frac{f(\hat{z}_k^+, \hat{z}_k^{+*}) + \epsilon_{k,+} - \left( f(\hat{z}_k^-, \hat{z}_k^{-*}) + \epsilon_{k,-} \right)}{2c_k \Delta_{k,i}^*},
	\label{eq:gradient}
\end{equation}
where
\begin{equation}
	\hat{z}_k^\pm = \hat{z}_k \pm c_k \Delta_k,
	\label{eq:perturb}
\end{equation}
$c_k$ is another positive gain coefficient, $\Delta_k$ is a random vector with entries independently chosen from $\{\pm 1, \pm i\}$, and $\epsilon_{k,\pm}$ represent possible measurement noise.

The coefficients $a_k$ and $c_k$ decay according to
\begin{equation}
	a_k = \frac{a}{(k+1+A)^s}, \quad c_k = \frac{b}{(k+1)^r},
	\label{eq:gain}
\end{equation}
where $a$, $b$, $A$, $s$, and $r$ are tuning parameters chosen to optimize the convergence rate. The standard choices are $a \sim 1$, $b \sim 0.1$, $s = 0.602$, $r = 0.101$, and $A \sim K/10$, with $K$ the number of iterations. However, these choices do not guarantee an optimal convergence rate, and they must be retuned for each problem to maximize performance.

CSPSA possesses fundamental properties that make it well-suited for variational quantum algorithms. First, it asymptotically converges in mean to a local optimum of the objective function. Second, the gradient estimator $\hat{g}_k$ is asymptotically unbiased, provided the errors $\epsilon_{k,\pm}$ have zero mean and are statistically independent across iterations. Third, evaluating the gradient estimator requires only two evaluations of the objective function, regardless of the system dimension or the number of parameters. This makes CSPSA well-suited for the experimental implementation of variational quantum algorithms, as it guarantees convergence in a sufficient number of iterations even in the presence of environmental and sampling noise.

\subsection{Classical shadows}

Classical shadows \cite{huang2020} provide an efficient strategy for evaluating the expectation values of many observables from a small number of single-shot quantum measurements. Here, we describe the local Pauli classical shadows.

Consider $N$ identically and independently prepared copies of an $n$-qubit quantum state $\rho$. For each copy, we apply a random unitary $U=\bigotimes_{j=1}^n U_j$, uniformly drawn from the single-qubit Clifford group. This group is generated by the gates $\{ \mathbb{I}, H, S^\dagger H\}$, with $\mathbb{I}$ the identity, $H$ the Hadamard gate, and $S$ the phase gate. This is equivalent to choosing, independently for each qubit, one of the three Pauli bases with equal probability. After that, we measure in the computational basis to obtain a binary outcome $\boldsymbol{b}=(b_1,\cdots,b_n)$. Depending on the random unitary $U$, this measurement corresponds to a projection onto the basis of the eigenvector of one of the Pauli matrices $\sigma_j$. The local-Pauli classical shadow of $\rho$ associated with $(U,\boldsymbol{b})$ is defined by
\begin{align}
    \hat{\rho}_{U,\boldsymbol{b}} = \bigotimes_{j=1}^n \left( 3 U_j^\dagger \dyad{b_i}{b_i} U_j - \mathbb{I} \right).
\end{align}
By construction, we have $\mathbb{E}_{U,\boldsymbol{b}}[\hat{\rho}_{U,\boldsymbol{b}}] = \rho$, so any linear function $\ev{O} = \Tr[O\rho]$ can be estimated as 
\begin{align}
    \ev*{\hat{O}} = \mathbb{E}_{U,\boldsymbol{b}}[ \Tr[O\hat\rho_{U,\boldsymbol{b}}] ]. \label{eq:shadow_observable}
\end{align}
The key feature is that we can avoid the explicit construction of the classical shadow $\hat{\rho}_{U, \boldsymbol{b}}$, which is an exponentially sized matrix, by directly computing $\Tr[O\hat\rho_{U,\boldsymbol{b}}]$. This allows for efficient estimation of any $k$-local observable $O$ with a sample complexity independent of the system size $n$, scaling as 
\begin{align}
    N =  \mathcal{O}(3^k \norm{O}^2_\infty/\epsilon^2) \label{eq:sample_complex}
\end{align}
for estimation error $\epsilon$. The factor $3^k$ in Eq.\thinspace\eqref{eq:sample_complex} grows exponentially with the number of qubits involved in the observable, so terms acting on all parties of the network are far the most expensive to resolve. 

\section{Method}

\subsection{Self-guided algorithm}
In this section, we describe the self-guided method for maximizing the nonlocal inequality in quantum networks. Let us consider a nonlocal inequality $I\leq I_{\rm Cl}$ for an arbitrary-size network. Let us suppose that the party $x$ performs measurements of dimension $d_x$. The self-guided algorithm is built from four key components:
\begin{enumerate}
    \item Complex parameterization of the measurement of each party $\{M_{a_x|\alpha_x}^{x}\}_{x\in\mathcal{X}}$. This parameterization must be expressive enough to capture the solution to the optimization problem, but excessive expressivity could lead to a barren plateau. The optimization parameters will be $z=\{z_x\}_{x\in\mathcal{X}}$. This assumption makes the optimization stage a device-dependent protocol.
    
    \item A set of $N_{\rm sh}$ Pauli shadows $\{\hat\rho_{U,b}\}$ is collected, which will be used to evaluate the inequality $I(z)$ on the parametrized measurement. This requires estimating a small set of observables $\{O_j\}$, one for each $k$-local term appearing in the inequality, which is done via Eq.\thinspace\eqref{eq:shadow_observable}. Since the number of such terms is small, the evaluation with Pauli shadows is efficient, provided $N_{\rm sh}$ is large enough to satisfy the sample-complexity bound of Eq.\thinspace\eqref{eq:sample_complex} for the desired estimation error $\epsilon$.

    \item Optimization by CSPSA. We consider an initial estimator $\hat{z}_0$. In each iteration $k$ of CSPSA with estimator $z_k$, two evaluations $I(z_k^\pm)$ are calculated from the set of classical shadows, with the perturbations $z_k^\pm$ given by Eq. \thinspace\eqref{eq:perturb}. These are used to estimate the gradient according to Eq.\thinspace\eqref{eq:gradient}. After that, we update the estimator to $z_{k+1}$ according to Eq.\thinspace\eqref{eq:update}. If required, the estimator $z_{k+1}$ is projected onto the set of feasible measurements. 

    \item Direct verification. Once the optimization has been run for $K$ iterations, the estimator $\hat{z}_K$ fixes a specific entangling measurement at each party. The value of $I(\hat z_K)$ used to drive the search is estimated only from classical shadows, a statistical, post-processed reconstruction that assumes a specific model of the measurement apparatus. This estimate alone cannot certify nonlocality, as imperfectly characterized tomographic reconstructions and entanglement witnesses are known to yield false positives \cite{rosset2012}. We therefore add a final stage in which the measurement defined by $\hat{z}_K$ is implemented directly at each party, and the probabilities entering $I(\hat{z}_K)$ are estimated from the observed outcome frequencies of this single, fixed measurement, over $N_{c}$ repetitions. If the value obtained in this way exceeds $I_{Cl}$ by a statistically significant margin, then the observed probability distribution is incompatible with Eq.\thinspace\eqref{eq:hidden_model}.
\end{enumerate}

We apply this self-guided algorithm to maximize the Wagon-Wheel inequality in the triangle network. In this case, three sources generate two-qubit states, so the global dimension of the system is $d=2^6$. Each party receives one qubit from each of the two sources it is connected to and performs a joint two-qubit measurement. Given this small local dimension, the measurement can be fully parametrized by an arbitrary four-dimensional unitary matrix followed by a computational-basis measurement without being strongly affected by the barren plateau phenomenon. We therefore parametrize the measurement of each party by a complex $d_x\times d_x$ matrix $Z_x$, which we project as $Z_x(Z_x^\dagger Z_x)^{-1/2}$ when unitarity has to be enforced. The evaluation of $I_{\rm WW}$ during the search requires estimating 15 distinct probabilities, far fewer than the $\mathcal{O}(d^2)$ settings needed for complete state tomography.

Note that the shadow dataset $\{\hat\rho_{U,b}\}$ is acquired once, before the search, and reused for every evaluation within it. This removes the need to implement a new measurement at each iteration, as other self-guided algorithms do, but it comes at a price. With a frozen dataset, the errors $\epsilon_{k,\pm}$ are neither zero-mean nor independent between iterations, as required for the convergence of CSPSA, since they are a fixed, deterministic function of the parameters $z$, set once and for all by the particular sample that was drawn. The optimizer is therefore free to maximize the sampling fluctuations of that specific dataset, driving the estimated value beyond the Fritz value $I_{\rm WW}^{\rm Fritz}$ while the underlying statistics need not violate the inequality at all. This bias makes the shadow-based value unusable as a certificate, which is why the direct verification is fundamental. The bias can be suppressed by using a number of shadows large enough that the estimation error in Eq.\thinspace\eqref{eq:sample_complex} is smaller than the gap between $I_{\rm Cl}$ and $I_{\rm WW}^{\rm Fritz}$, so that the classical bound and the maximal violation can be resolved from one another.

\begin{algorithm}[h!]
\caption{Self-guided maximization of $I_{\rm WW}$.}

\KwIn{Three two-qubit states $\rho_{AB}, \rho_{BC}, \rho_{CA}$,
number of shadows $N_{\rm sh}$, number of verification repetitions $N_{\rm c}$, CSPSA hyperparameters $\{a_k, c_k\}$,
initial complex $4\times 4$ matrices $Z_{A,0}, Z_{B,0}, Z_{C,0}$.}
Prepare $\rho = \rho_{AB} \otimes \rho_{BC} \otimes \rho_{CA}$.\\
Collect $N_{\rm sh}$ Pauli shadows $\{\hat{\rho}_{U,b}\}$ of $\rho$.\\
Set $\hat{z}_0 = \{\hat{Z}_{x,0} \}$ with $x\in\{A,B,C\}$.\\
\For{$k = 0, \dots, K $}{
    Sample perturbation vector $\Delta_k$.\\
    Compute $\hat{z}_k^+$ and $\hat{z}_k^-$\\
    Project $\hat{Z}_{x,k}^\pm \leftarrow \hat{Z}_{x,k}^\pm (\hat{Z}_{x,k}^{\pm\dagger} \hat{Z}_{x,k}^\pm)^{-1/2}$.\\
    Evaluate $I_{\rm WW}( \hat{z}_k^+)$ and $I_{\rm WW}(\hat{z}_k^-)$ from $\{\hat{\rho}_{U,b}\}$. \\
    Estimate gradient $\hat{g}_k$.\\
    Update $\hat{z}_{k+1} \leftarrow \hat{z}_k + a_k \hat{g}_k$\\
    Project $\hat{Z}_{x,k+1} \leftarrow \hat{Z}_{x,k+1} ( \hat{Z}_{x,k+1}^\dagger \hat{Z}_{x,k+1})^{-1/2}$.\\
    Evaluate $I_{\rm WW}(\hat{z}_{k+1})$ from $\{\hat{\rho}_{U,b}\}$.\\ 
}
Fix the measurement $\hat{Z}_{x,K}$ at each party $x$.\\
Evaluate $I_{\rm WW}(\hat{z}_K)$ from $N_{\rm c}$ measurement of $\hat{Z}_{x,K}$.\\
\KwRet{$I_{\rm WW}(\hat{z}_K)$}.
\end{algorithm}

\section{Results}

We numerically simulate the self-guided algorithm that maximizes the Wagon-Wheel inequality $I_{\rm WW}$ in the triangle network, estimating $I_{\rm WW}$ at each iteration from a finite set of $N_{\rm sh}$ classical shadows. The CSPSA hyperparameters used in all simulations are $s=0.602$, $r=0.101$, $A=0$, $a=3$, and $b=0.1$. The initialized local basis $\hat{Z}_{x,0}$ is uniformly drawn from a Haar random distribution. We consider three families of two-qubit states generated by each source. First, the maximally entangled singlet state 
\begin{align}
    \ket{\psi^-}=\frac{1}{\sqrt{2}}(\ket{01}-\ket{10})
\end{align}
Second, Schmidt states
\begin{align}
    \ket{\psi(a)} = \sqrt{1-a}\,\ket{00} + \sqrt{a}\,\ket{11}, \quad a\in[0,1/2],
    \label{eq:schmidt}
\end{align}
which are pure but only partially entangled for $a<1/2$. Finally, Werner states
\begin{align}
    \rho(p) = p\,\ket{\psi^-}\!\bra{\psi^-} + (1-p)\,\frac{\mathbb{I}_4}{4}, \quad p\in[0,1],
    \label{eq:werner}
\end{align}
an isotropic mixture of the singlet with white noise, entangled for $p>1/3$ and CHSH-nonlocal for $p>1/\sqrt{2}$. Schmidt states model coherent noise, that is, pump or interferometer imbalance in entangled-photon sources, which preserves purity while reducing entanglement. Werner states model incoherent noise, that is, photon loss or depolarization in long-distance channels, which instead reduces purity.
 
For each of these families, we repeat the simulation $10^2$ times and report the typical performance of the protocol as the median, together with the interquartile range. We consider that our protocol certifies nonlocality when the entire interquartile range lies above the classical bound $I_{\rm Cl}$, which means that at least $75\%$ of the realizations violate the inequality and indicate nonlocality. This criterion is a statement about the typical behavior of the protocol over repeated runs of the algorithm.

\begin{figure*}[t]
\centering
\includegraphics[width=0.95\textwidth]{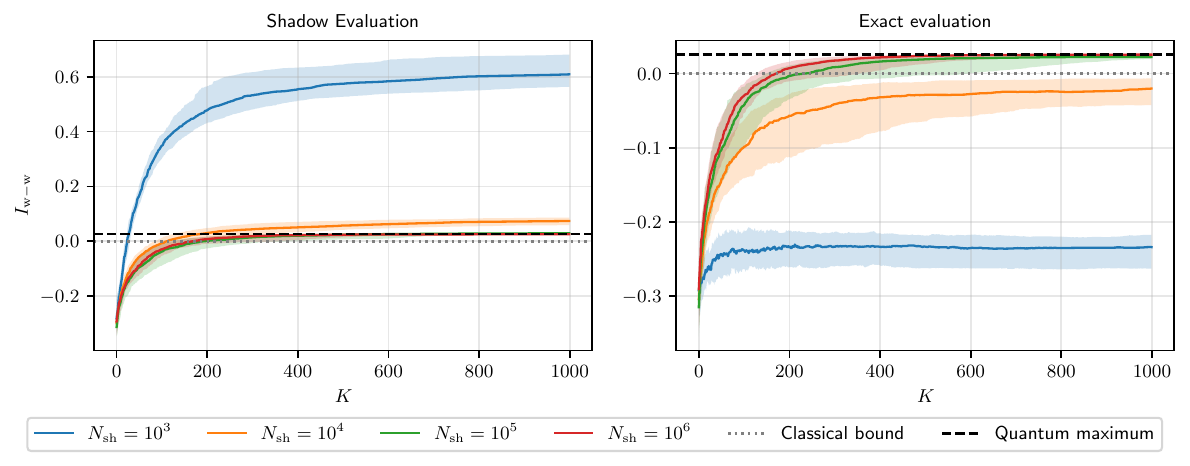}
\caption{Wagon-Wheel inequality $I_{\rm WW}$ as a function of the iteration number $k$, for the three sources emitting the maximally entangled state $\ket{\psi^-}$ and several numbers of classical shadows $N_{\rm sh}$. Solid lines show the median over $10^2$ realizations and shaded regions show the interquartile range. Left: value estimated from $N_{\rm sh}$ shadows, used to drive the CSPSA optimization. Right: exact value of $I_{\rm WW}$ at the same measurement settings. Dotted and dashed lines indicate the classical bound and the Fritz value, respectively.}
\label{fig:iterations}
\end{figure*}
 
Figure~\ref{fig:iterations} shows, for the three sources that generate and emit the maximally entangled states $\ket{\psi^-}$, the median of $I_{\rm WW}$ that drives the optimization (left, estimated from $N_{\rm sh}$ shadows) together with its exact value at the same measurement settings (right), as a function of the iteration number $k$, for $N_{\rm sh}\in\{10^3,10^4,10^5,10^6\}$. The shaded regions show the interquartile range over the $10^2$ realizations. The right panel corresponds to the noise-free limit of the direct verification step (step 4). In an actual implementation, this value would instead be obtained by directly measuring the fixed and optimized measurement in each party, so a genuine violation of $I_{\rm Cl}$ certifies nonlocality independently of any bias introduced by the shadow-based search. For $N_{\rm sh}=10^3$ the shadow-based estimate grows to a median $I_{\rm WW}\approx0.6$, more than twenty times the quantum maximum $I_{\rm WW}^{\rm Fritz}\approx0.025$, while the exact value in those same settings has a median around $I_{\rm WW}\approx-0.25$, with its entire interquartile range well below the classical bound $I_{\rm Cl}=0$. This happens because the optimizer is driven entirely by the statistical fluctuations of the shadow dataset, and the measurement setting it returns does not detect nonlocality at all. For $N_{\rm sh}=10^4$ the bias is smaller but still present, and the median exact value approaches $I_{\rm WW}\approx-0.02$ by $K=10^3$. However, for more iterations, the entire interquartile range exceeds the Fritz value, indicating that the optimization is also biased by the shadow uncertainty. This suggests that even with a certification of nonlocality at low iterations, we must run many iterations to ensure that the optimization bias is small. Only for $N_{\rm sh}=10^5$ and $10^6$ does the interquartile range of the exact value fully cross the classical bound, at $K\approx250$ and $K\approx180$, respectively, after which both the median and the interquartile range converge to the Fritz value. The magnitude of these fluctuations is consistent with the sample-complexity bound for classical shadows, Eq.~\eqref{eq:sample_complex}. The most demanding terms of $I_{\rm WW}$ are joint correlators over all six qubits of the network, that is, $k=6$, for which resolving a violation of order $I_{\rm WW}^{\rm Fritz}\approx0.025$ requires an estimation error $\epsilon\lesssim I_{\rm WW}^{\rm Fritz}$, that is, $N\gtrsim 3^6/0.025^2\approx1.2\times10^6$ shadows, in good agreement with the shadow budget at which the protocol becomes reliable in Fig.~\ref{fig:iterations}. Since this estimate follows from a worst-case bound, it also explains why $N_{\rm sh}=10^5$, although below this threshold, already comes close to the Fritz value. Thus, a sufficiently large number of iterations and a shadow budget are required for the protocol to reliably certify nonlocality.

\begin{figure*}[t]
\centering
\includegraphics[width=0.95\textwidth]{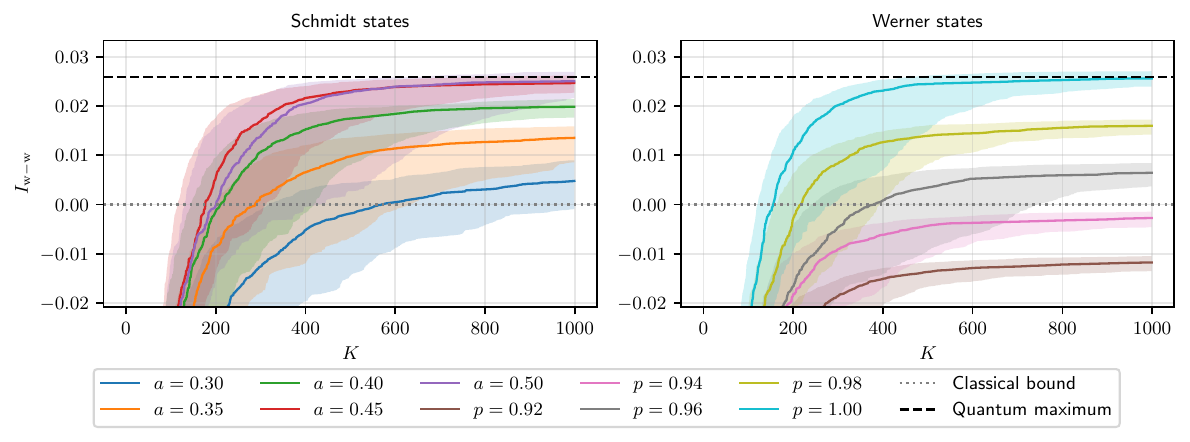}
\caption{Wagon-Wheel inequality $I_{\rm WW}$ as a function of the iteration number $k$, for $N_{\rm sh}=10^6$. Solid lines show the median over $10^2$ realizations and shaded regions show the interquartile range. Left: Schmidt states with entanglement parameter $a$. Right: Werner states with mixing parameter $p$. Dotted and dashed lines indicate the classical bound and the Fritz value, respectively.}
\label{fig:states}
\end{figure*}

Figure~\ref{fig:states} shows the median of $I_{\rm WW}$ reached by the protocol, together with its interquartile range (shaded), as a function of the iteration number, for Schmidt states with $a\in\{0.30,0.35,0.40,0.45,0.50\}$ (left) and Werner states with $p\in\{0.92,0.94,0.96,0.98,1.00\}$ (right), using $N_{\rm sh}=10^6$. For Schmidt states, the interquartile range fully crosses the classical bound for $a\geq0.35$, doing so earliest for the most entangled states ($a=0.45,0.50$, at $K\approx250$) and last for $a=0.35$ (at $K\approx600$). For the least entangled state, $a=0.30$, the median crosses the classical bound around $K\approx560$, but the interquartile range never fully clears it within $10^3$ iterations, so this state is not certified by our criterion within the simulated range. For Werner states, the interquartile range fully crosses the classical bound for $p=1.00$ ($K\approx200$), $p=0.98$ ($K\approx600$) and $p=0.96$ ($K\approx750$), while for $p=0.94$ the entire interquartile range, including its upper edge, remains below the classical bound throughout $10^3$ iterations, and for $p=0.92$ it is even worse. Overall, Schmidt and Werner state that maximal violations farther from the classical bound are easier to certify, and that the number of iterations required grows as the target violation shrinks.

We can compare our Werner threshold with the value $p\approx0.94$ reported in Ref.\thinspace\cite{sunilkumar2025} as the best numerical noise-robustness bound for genuine nonlocality of Werner states in the triangle network. We can see that we were able to certify nonlocality just above $p\approx0.94$, so our results are consistent with this previously known threshold.

\section{Conclusions}

We introduced a self-guided protocol to certify nonlocality in quantum networks, in which the local measurement settings of each party are variationally optimized to maximize the violation of a network Bell inequality. The violation is estimated at each iteration from a finite set of local Pauli classical shadows, and the device-dependent search is driven by CSPSA. After convergence, we implement the optimized measurement directly at each party and estimate the inequality from the observed outcome frequencies rather than from classical shadows. Because classical shadows are post-processed statistical estimators that a local hidden-variable model could, in principle, reproduce even when it indicates a violation \cite{rosset2012}, this final direct evaluation is what actually certifies nonlocality. Applied to the triangle network and the Wagon-Wheel inequality, the protocol converges toward the theoretically predicted quantum maximum for the Fritz distribution, provided that a sufficiently large shadow budget and a sufficiently large number of iterations are used. 

The optimization step and the direct certification differ in their assumptions. The search is device-dependent, since the shadow estimator inverts an assumed model of the measurement apparatus \cite{huang2020}, but its only role is to propose settings. What it returns is a recipe, that is, an inequality together with a fixed measurement at each party. A user who receives only this recipe, with no knowledge of how it was obtained, evaluates the inequality on the raw outcome frequencies of black boxes, and a violation then certifies nonlocality device-independently \cite{acin2007,brunner2014}. The protocol is thus device-dependent, yet its output defines a device-independent test.

Our simulations show that the performance of the protocol as a nonlocality certification is jointly governed by the shadow budget $N_{\rm sh}$, which sets the bias of the estimated objective relative to its exact value, and the number of iterations $K$ needed for convergence to optimal measurement settings. A shadow budget below the sample-complexity threshold set by the locality of the inequality's most demanding terms can produce an artificially large violation while the true state remains uncertified, so $N_{\rm sh}$ and $K$ must be chosen jointly, with both growing as the target violation shrinks toward the classical bound. We showed that the protocol correctly certifies both partially entangled Schmidt and noisy Werner states, being especially efficient when their entanglement is high.

Semidefinite programming cannot be applied directly to nonlinear inequalities in quantum networks, since it requires a linear objective function \cite{tavakoli2024}. Solving them via SDP therefore requires first relaxing the nonlinear problem into a convergent hierarchy of semidefinite programs \cite{renou2026,klep2023}. Our protocol instead optimizes the nonlinear objective directly, requiring only the ability to perform parametric measurements at each party and experimentally evaluate the inequality. 

Beyond the triangle network, our protocol applies to any nonlinear network inequality that can be expressed in terms of local Pauli observables. In larger systems, working with highly local inequalities becomes crucial. When a term involves a global observable that acts on all parties, the classical-shadow estimator experiences the phenomenon of barren plateaus, which requires an exponential number of shots to resolve \cite{cerezo2021,arrasmith2021, Zambrano2024}. Our protocol should therefore work best for inequalities built from few-body terms, which motivates the search for such inequalities. Numerical approaches already exist for constructing Bell inequalities from few-body correlators in many-body systems \cite{tura2014,frerot2021}, and extending them to the network setting is a promising direction. A complementary route to a speed-up is to restrict the optimization to a specific parametric family of measurements rather than an arbitrary unitary at each party. Here we exploited the small local dimension to optimize over arbitrary measurements, but for larger systems, departing from a unitary 2-design in favor of a physically motivated, less expressive ansatz could improve trainability \cite{cerezo2021}.  

A natural next step is an experimental demonstration of the protocol on a real quantum network, and several existing photonic realizations of the triangle network are well suited for this. Suprano \textit{et al.} \cite{suprano2022} implemented a triangle network with three independent polarization-entangled photon-pair sources, certifying genuine tripartite nonlocality through separable measurements and chained Bell inequalities. Meskine \textit{et al.} \cite{meskine2025} simulated a fiber-based telecom triangle network from a single multiplexed AlGaAs source, explicitly modeling and violating a noise-robust Bell-like inequality. Wang \textit{et al.} \cite{wang2026} realized a six-photon triangle network that produces the Elegant distribution and certified its nonlocality. In these experiments, the measurement settings were fixed in advance or tuned via classical or machine-learning-assisted searches, rather than being variationally optimized from the observed statistics. Our self-guided protocol could be run directly on such setups, using measured classical shadows to drive the local measurement bases in real time, providing a practical, minimal-assumption tool for certifying nonlocality under realistic experimental noise. Beyond photonics, distributed quantum computing platforms are a further natural setting. Trapped-ion modules connected by an optical link have already been used to distribute computations between separated processors \cite{main2025}, and superconducting circuits joined by a cryogenic link have violated a CHSH inequality loophole-free over $30$ m \cite{storz2023}. Neither platform has yet realized a triangle network with three independent sources, but both provide the entangling measurements and source independence that the protocol requires, so such a realization appears feasible in the near term.

An advantage of our method on a photonic platform is that training requires only local single-qubit measurements, never an entangling one. This has a fundamental benefit as it avoids the inherently probabilistic nature of entangling photons via linear-optical operations. For example, a linear-optical Bell state analyzer can unambiguously distinguish only two of the four Bell states, so its success probability cannot exceed $50\%$ \cite{calsamiglia2001}, and implementing a photonic CNOT gate with linear optics succeeds with probability as low as $1/9$ \cite{knill2001,obrien2003}. In our protocol, only the final direct verification step requires the actual entangling measurement fixed by the optimization, so this limitation affects only a single measurement at the end of the protocol rather than every evaluation performed during the search.

\begin{acknowledgments}

J.C.-V. and A.D. were supported by the National Agency of Research and Development (ANID) -- Millennium Science Initiative Program -- ICN17$_-$012. A.D. acknowledges financial support from FONDECYT Regular Grant No. 1230586. J.C.-V. was supported by CONICYT-PCHA / DoctoradoNacional/2018-21181692. L.P. was supported by the Government of Spain (Severo Ochoa CEX2019-000910-S, FUNQIP, and QEC4QEA PCI2025-163167), European Union (PASQuanS2.1, 101113690, QSNP, 101114043, and QEC4QEA, 101194322), Fundació Cellex, Fundació Mir-Puig, and Generalitat de Catalunya (CERCA program). 

\end{acknowledgments}

\end{document}